\documentclass[letter,traditabstract]{aa}

\usepackage{graphicx}
\usepackage{txfonts}
\usepackage{natbib}
\bibpunct{(}{)}{;}{a}{}{,}

\def\nul#1{}
\def\be \begin{equation}
\def\ee \end{equation}
\def\figpath{}

\def\Sa{{\cal S}_1}
\def\Sb{{\cal S}_2}

\begin{document}

\title{{HD}\,60532,  a planetary system in a 3:1 mean motion resonance.}
\titlerunning{{HD}\,60532,  a planetary system in a 3:1 mean motion resonance}

\author{
J.~Laskar\inst{1} 
\and
A.C.M. Correia\inst{1,2}
}

\offprints{J. Laskar}
 
\institute{
  ASD, IMCCE, CNRS-UMR8028, Observatoire de Paris, UPMC, 77 avenue
  Denfert-Rochereau, 75014 Paris, France \\
  \email{laskar@imcce.fr}
  \and
Departamento de F\'isica, Universidade de Aveiro, Campus de
  Santiago, 3810-193 Aveiro, Portugal \\
  \email{correia@ua.pt}
}

\date{Received ; accepted To be inserted later}

  \abstract{In a recent paper it was reported a planetary
  system around the star HD\,60532, composed by two giant planets in a possible
  3:1 mean motion resonance, that should be confirmed within the next decade.
  Here we show that the analysis of the global dynamics of the system
allows to confirm this resonance. The present best
  fit to data already corresponds to this resonant configuration and the system
  is stable for at least 5\,Gry. The 3:1 resonance is so robust that stability
  is still possible for a wide variety of orbital parameters around the best fit
  solution and also if the inclination of the system orbital plane with respect
  to the plane of the sky is as small as $15^\circ$. Moreover, if the
  inclination is taken as a free parameter in the adjustment to the
  observations, we find an inclination $ \sim 20^\circ $, which corresponds to $M_b =
  3.1\,M_{\rm Jup}$ and $ M_c = 7.4\,M_{\rm Jup}$ for the planetary companions.}

   \keywords{stars: individual: {HD}\,60532 -- stars: planetary systems --
   techniques: radial velocities }

   \maketitle
%


\section{Introduction}

In a recent paper \citep[hereafter DES08]{Desort_etal_2008}, two planetary
mass companions were detected around the F type star {\small HD}\,60532.
Using the data acquired with the HARPS spectrograph based on the 3.6-m ESO
telescope at La Silla Observatory, and a two-keplerian fit to the data, 
the orbits of the two bodies were determined, corresponding to
minimum masses of 1.03\,$M_{Jup}$ and 2.46\,$M_{Jup}$ at 201-day and 604-day
period, respectively.
The dynamical study done in DES08
suggested a possible 3:1 resonance, but the stability of the results seemed
to be questionable, as small variations of the semi major axis 
led to a non resonant solution (DES08). 
Therefore, the authors   concluded that the
existence of a mean motion resonance would require an additional decade 
of observations to be confirmed.

The presence of two or more interacting planets in a system dramatically
increases our potential ability to constrain and understand the processes of
planetary formation and evolution. The dynamical analysis of such systems is
then very useful, first for constraining the system evolution history and second
to determine the system ``structure'' in terms of orbital content.  
Among the known multi-planet systems, a significant fraction present strong
interactions and are trapped in mean motion resonances.
The resonances between planets are believed to be
formed after their inward or outward migration of the planets during
the early stages of the formation of the system 
\citep[eg.][]{Tsiganis_etal_2005}.

By performing a detailed dynamical analysis of the HD\,60532 system in
conformity with the radial-velocity observations, we can   confirm
this system in a 3:1 mean-motion resonance and provide some constraints on the
inclination of the system orbital plane with respect to the plane of the sky.
In Section\,2 we re-analyze the observational data obtained with the {\small
HARPS} spectrograph and the dynamical analysis of the system discussed in
Sect.\,3. Finally, conclusions are drawn in Sect.\,4.


\section{Orbital solution for the HD\,60532 system}
\label{orbsolutions}

The 147 published radial-velocity data points of HD\,60532 (DES08) were taken
with the {\small HARPS} spectrograph from February 2006 until June 2008. 
A two-companion Keplerian solution is provided in DES08, but because of the
strong mutual interactions between the two planets, these parameters present
important variations  and cannot
be used in dynamical studies unless   the
initial longitudes of
both planets are given.
Using the iterative Levenberg-Marquardt method \citep{Press_etal_1992}, we thus 
re-fitted
the complete set of radial velocities with a 3-body Newtonian model, assuming
co-planar motion perpendicular to the plane of the sky, similarly to 
what has been done for the system {\small HD}\,45364 \citep{Correia_etal_2009}.
This fit yields an adjustment of $\sqrt{\chi^2}=4.41$ and $rms=4.38\,\mathrm{m
s}^{-1}$, slightly better than the two-Keplerian model from DES08. 
The set of orbital parameters for this system ($\Sa$) is given in Table\,\ref{T1}.

Still assuming co-planar motion, we then release the perpendicularity constraint 
by including the inclination as a free parameter.
The system readjusts slightly the orbital parameters of the two
planets and provides for the best stable fit an inclination around $20^\circ$,
although subjected to a large uncertitude. 
Fixing the inclination at this value the new fit yields an adjustment of
$\sqrt{\chi^2}=4.37$ and $rms=4.34\,\mathrm{m s}^{-1}$.
In this case the masses are increased by a factor $ 1/\sin i = 2.92 $, i.e., $M_b
= 3.1\,M_{\rm Jup}$ and $ M_c = 7.4\,M_{\rm Jup}$ ($\Sb$, Table\,\ref{T1}).
We also attempted to fit the data with a 3-body Newtonian model for which the
inclination of the orbital planes, as well as the node of the outer planet
orbit, was free to vary.  
We were able to find a wide variety of configurations, some with very low
inclination values for one or both planets, that slightly improved the fit.
However, all of these determinations are uncertain, and since we also increased
the number of free parameters by three, we cannot say that these solutions
present  a real improvement with respect to the sets $\Sa,\Sb$, presented in
Table\,\ref{T1}.

\begin{table}
\caption{Orbital parameters for the HD\,60532 system $\Sa$, obtained
with a 3-body Newtonian fit with $i=90^\circ$ (top) 
and for the system system $\Sb$, with $i=20^\circ$ (bottom) ($M_\star = 1.44 \,M_\odot$).  \label{T1}}
\begin{center}
\begin{tabular}{l l c c} \hline \hline
\noalign{\smallskip}
{\bf Param.}  & {\bf [unit]} & {\bf {HD}\,60532\,b} & {\bf {HD}\,60532\,c} \\ \hline 
\noalign{\smallskip}
$\sqrt{\chi^2}$     &                      & \multicolumn{2}{c}{4.407}   \\ 
$rms$        & [m/s]                & \multicolumn{2}{c}{4.380}   \\  \hline
\noalign{\smallskip}
Date         & [JD-2400000]         & \multicolumn{2}{c}{54000.00 (fixed)}  \\ 
$V$          & [km/s]               & \multicolumn{2}{c}{$ -0.0049 \pm 0.0004 $}  \\  
$P$          & [day]                & $ 201.46 \pm 0.13  $ & $ 605.28 \pm 2.12  $ \\ 
$\lambda$    & [deg]                & $  14.48 \pm 0.69  $ & $ 316.23 \pm 0.95  $ \\ 
$e$          &                      & $  0.279 \pm 0.006 $ & $  0.027 \pm 0.007 $ \\ 
$\omega$     & [deg]                & $ 352.15 \pm 1.08  $ & $ 136.81 \pm 16.34  $ \\ 
$K$          & [m/s]                & $  29.63 \pm 0.32  $ & $  46.80 \pm 0.40  $ \\  
$i$          & [deg]                & $ 90 $ (fixed)      & $ 90 $ (fixed)      \\  \hline
\noalign{\smallskip}
$M  $ & [M$_\mathsf{Jup}$]     & $ 1.0484 $           & $ 2.4866 $ \\
$a$          & [AU]                 & $ 0.7597 $           & $ 1.5822 $ \\ 
\noalign{\smallskip}
\hline
\hline 
\noalign{\smallskip}
{\bf Param.} Ê& {\bf [unit]} & {\bf {HD}\,60532\,b} & {\bf
{HD}\,60532\,c} \\ \hline
\noalign{\smallskip}
$\sqrt{\chi^2}$     &                      & \multicolumn{2}{c}{4.369}   \\
$rms$        & [m/s]                & \multicolumn{2}{c}{4.342}   \\ 
\hline
\noalign{\smallskip}
Date         & [JD-2400000]         & \multicolumn{2}{c}{54000.00 (fixed)}  \\
$V$          & [km/s]               & \multicolumn{2}{c}{$ -0.0055    \pm 0.0003 $}  \\
$P$          & [day]                & $ 201.83 \pm 0.14  $ & $ 607.06 \pm 2.07  $ \\
$\lambda$    & [deg]                & $  14.78 \pm 0.66  $ & $ 317.02 \pm 0.93  $ \\
$e$          &                      & $  0.278 \pm 0.006 $ & $  0.038 \pm 0.008 $ \\
$\omega$     & [deg]                & $ 352.83 \pm 1.05  $ & $ 119.49 \pm 9.14  $ \\
$K$          & [m/s]                & $  30.34 \pm 0.32  $ & $  47.84 \pm 0.44  $ \\
$i$          & [deg]                & $ 20 $ (fixed)      & $ 20 $(fixed) \\  \hline
\noalign{\smallskip}
$M$          & [M$_\mathsf{Jup}$]   & $ 3.1548 $           & $ 7.4634 $ \\
$a$          & [AU]                 & $ 0.7606 $           & $ 1.5854
$ \\ \hline
\end{tabular}
\end{center}
\end{table}
\section{The 3:1 mean motion resonance}
\label{31res}

\begin{figure}
  \centering
    \includegraphics*[width=7.cm]{\figpath 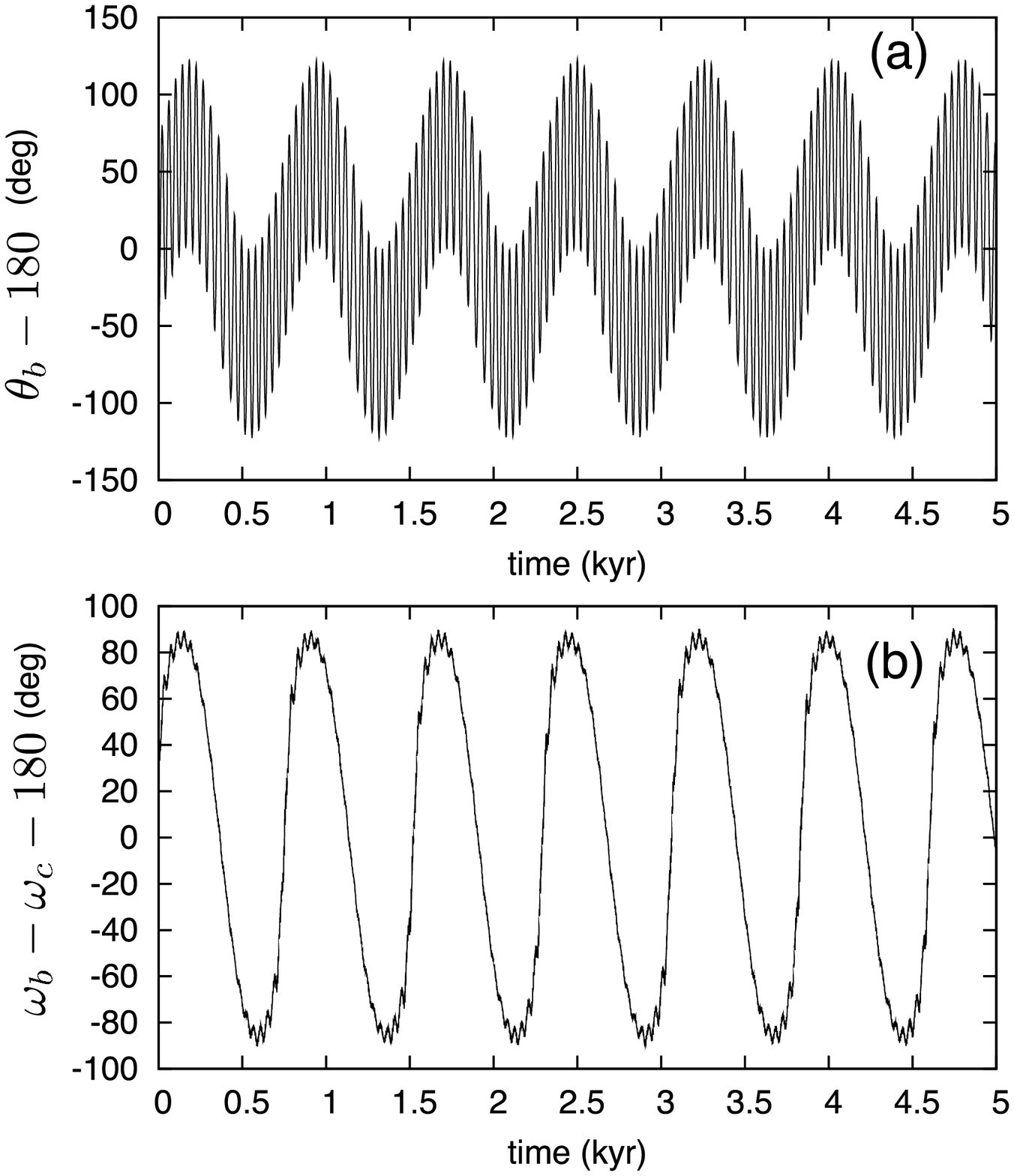} 
  \caption{Variation in the resonant argument, $\theta =  \lambda_b - 3
  \lambda_c + 2 \omega_b $ (a) and in the secular  argument, $ \Delta \omega =
  \omega_b - \omega_c $ (b), with time. \label{F3}}   
\end{figure}

\begin{figure*}
  \centering
    \includegraphics*[width=17cm]{\figpath 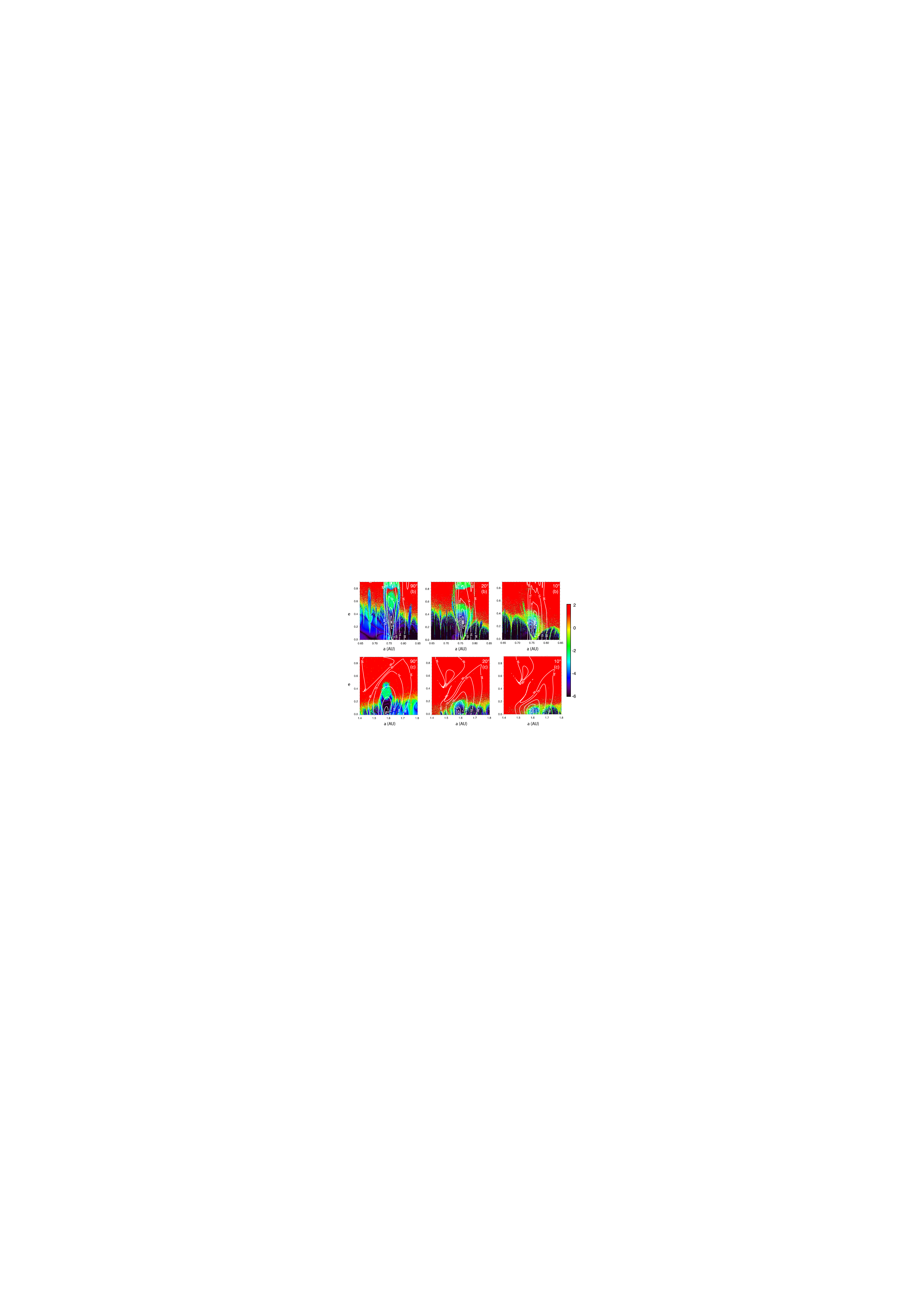} 
  \caption{
  Stability analysis  
  of the {HD}\,60532 planetary system for different values of the inclination ($90^\circ, 20^\circ, 10^\circ$). 
  For a fixed orbital inclination, and initial condition 
  of the outer  and inner planet, the phase space of the 
  system is explored by varying the semi-major axis $a_k$ and eccentricity 
  $e_k$ of the other planet, (b) and (c) respectively. The step size is $0.002$ AU in semi-major axis 
  and $0.005$ in eccentricity. For each initial condition, the 
  full system is integrated numerically over 10~kyr and, 
  as in \citep{Correia_etal_2005,Correia_etal_2009}, a stability criterion 
  is derived  with the frequency analysis of the mean longitude
  \citep{Laskar_1990,Laskar_1993PD}.
  The ``red'' zone corresponds to highly unstable 
  orbits, while  the ``dark blue'' region can be assumed to be stable on a
  billion-years timescale.
  The contour curves indicate the value of $rms$ 
  obtained for each choice of parameters.
  It is remarkable that in the present fit, there is perfect 
  correspondence between the  zone of minimal $rms$  
  and the 3:1 stable resonant zone, in ``dark blue''.
  \label{F4}}   
\end{figure*}

\begin{table}
 \caption{Fundamental frequencies for the orbital solution $\Sa$  (Tab.\ref{T1}).
$n_b$ and $n_c$ are the mean motions, $g_1$ and $g_2$ are the
 secular frequencies of the periastrons, and $l_\theta$ is the libration
 frequency of the resonant angle $\theta =  \lambda_b - 3 \lambda_c +
 2\omega_b $. 
 \label{T2}} 
 \begin{center}
 \begin{tabular}{crr}
 \hline\hline
      & Frequency   & Period \\
      & $^\circ/yr$ & yr \\
 \hline
 $n_b$ &   653.645983    &       0.550757\\
 $n_c$ &   217.563988    &       1.654686\\
 $g_1$ &    -0.477010    &    -754.701\\
 $g_2$ &    -0.009651    &   -37303.55\\
 $l_\theta$ & 9.015796    &      39.9299 \\ \hline
\end{tabular}
\end{center}
\end{table}


\begin{table}
\caption{Quasi-periodic decomposition of the resonant angle $\theta =  \lambda_b - 3
  \lambda_c + 2 \omega_b $ for  the orbital solution  $\Sa$ (Tab.\ref{T1}). 
    We have $\theta = 180+\sum_{i=1}^N A_i \cos(\nu_i\, t + \phi_i)$, where 
  the amplitude and phases $A_i$, $\phi_i$, are given in degree, and the 
  frequencies $\nu_i$ in degree/year. 
  All terms are identified as  integer combinations 
  of the fundamental frequencies given in
  Table\,\ref{T2}, which is a signature of a very regular motion. 
\label{T3}} 
  \begin{center}
    \begin{tabular}{rrrrrrrr}
\hline\hline
   \multicolumn{5}{c}{\textbf{Combination}} & \multicolumn{1}{c}{$\nu_i$} 
    &\multicolumn{1}{c}{$A_i$}  & \multicolumn{1}{c}{$\phi_i$}  \\
\multicolumn{1}{c}{$n_b$}  & \multicolumn{1}{c}{$n_c$}& 
 \multicolumn{1}{c}{$g_1$}  & \multicolumn{1}{c}{$g_2$} 
    &\multicolumn{1}{c}{ $l_\theta$}     & \multicolumn{1}{c}{(deg/yr)}
    &\multicolumn{1}{c}{(deg)}                & \multicolumn{1}{c}{(deg)}  \\
\hline
 0 &  0 & -1 &  1 &  0 &     0.4674 &        63.652 &      -81.114 \\ 
 0 &  0 &  0 &  0 &  1 &     9.0158 &        38.138 &      168.018 \\ 
 0 &  0 &  1 & -1 &  1 &     8.5484 &        35.869 &      159.131 \\ 
 0 &  0 &  2 & -2 &  1 &     8.0811 &        25.845 &      -29.755 \\ 
 0 &  0 &  3 & -3 &  1 &     7.6137 &        16.868 &      141.358 \\ 
 0 &  0 &  4 & -4 &  1 &     7.1464 &        10.622 &      -47.528 \\ 
 0 &  0 &  5 & -5 &  1 &     6.6790 &         6.325 &      123.585 \\ 
 0 &  0 &  6 & -6 &  1 &     6.2116 &         3.509 &      -65.301 \\ 
 0 &  0 & -2 &  2 &  0 &     0.9347 &         2.697 &      107.773 \\ 
 0 &  0 & -1 &  1 &  1 &     9.4832 &         2.366 &      176.904 \\ 
 0 &  0 &  7 & -7 &  1 &     5.7443 &         1.872 &      105.812 \\ 
 0 &  0 &  3 & -3 &  2 &    16.6295 &         1.156 &       39.376 \\ 
 0 &  0 &  2 & -2 &  2 &    17.0969 &         0.874 &     -131.738 \\ 
 0 &  0 & -2 &  2 &  1 &     9.9505 &         0.874 &        5.791 \\ 
 0 &  0 &  0 &  0 &  2 &    18.0316 &         0.828 &       66.035 \\ 
 0 &  0 &  8 & -8 &  1 &     5.2769 &         0.964 &      -83.074 \\ 
 0 &  0 &  4 & -4 &  2 &    16.1622 &         0.805 &     -149.510 \\ 
 0 &  0 &  1 & -1 &  2 &    17.5642 &         0.499 &     -122.851 \\ 
 0 &  0 & -3 &  3 &  1 &    10.4179 &         0.517 &     -165.323 \\ 
 0 &  0 & -3 &  3 &  0 &     1.4021 &         0.433 &      -63.341 \\ 
 0 &  0 &  9 & -9 &  1 &     4.8096 &         0.482 &       88.040 \\ 
 1 & -1 &  0 &  0 &  0 &   436.0820 &         0.395 &     -173.181 \\ 
  \hline
\end{tabular}
\end{center}
\end{table}

As in DES08, our nominal solution $\Sa$ (Tab.\ref{T1}) seems to be in a 3:1 resonance.
We integrated numerically the orbits of the planets  over 100 kyr with the 
symplectic integrator SABAC4 of \citet{Laskar_Robutel_2001}, using a step size
of 0.02~years. The frequency analysis of this 
orbital solution  allows then to  conclude that it is  indeed in a
3:1 mean motion resonance, with resonant argument:
$
\theta = \lambda_b - 3 \lambda_c + 2 \omega_b
$.

The fundamental frequencies of the systems are 
the two mean motions $n_b$ and $n_c$, the two secular frequencies of the 
periastrons $g_1$ and $g_2$, and the libration frequency of the resonant argument
$l_{\theta}$ (Tab.\ref{T3}). We have, up to the precision of the determination of the frequencies ($\approx 10^{-9}$), 
the resonant relation $n_b-3 n_c + 2 g_1 = 0 $.

For the nominal solution $\Sa$, the resonant argument $ \theta $ is in libration around $180^\circ $,
with a libration period $ 2 \pi / l_\theta = 39.93 $~yr, and an associated 
amplitude of about 38.138 degrees, with
multiples additional terms of significant amplitude of the form 
$k(g1-g2) +l_\theta$, with $k$ integer (Fig.\,\ref{F3}a, Table\,\ref{T3}). 
For the complete solution, the libration amplitude can reach
123.16 degrees because additional periodic terms are present. 
In Table\,\ref{T3}, we provide a quasi-periodic  decomposition 
of the resonant angle $ \theta $
in terms of decreasing amplitude.
All the quasi-periodic terms are easily identified as integer combinations of the 
fundamental frequencies (Table\,\ref{T3}).
Since the resonant angle is modulated by a relatively short period of about 38~years, 
and high harmonics of shorter periods, the
observation of the system over a few additional years may provide an
estimate of the libration amplitude and thus a strong constraint
on the parameters of the system.

As in (Correia et al 2009), 
both periastrons precess with mean 
frequency $g_1$ that is retrograde, with a period of 754.701 years. The two
periastrons are thus 
locked in an antipodal state, and the difference $\Delta \omega = \omega_b-\omega_c$ is in libration 
around $180^\circ$ with an amplitude of $90\fdg7$ (Fig.\,\ref{F3}b).
As a result, the arguments  $\theta' =  \lambda_b - 3 \lambda_c + 2\omega_c$,
and  $\theta'' =  \lambda_b - 3 \lambda_c +  \omega_b+  \omega_c$
librate around $180^\circ$ and $0^\circ$ respectively,  with the same libration frequency
$l_{\theta}$.

\section{Stability analysis}

To analyze the stability of the nominal solution and
confirm the presence of the 3:1 resonance, 
we performed a global frequency analysis \citep{Laskar_1993PD} 
in the vicinity of the nominal solution (Fig.\,\ref{F4}), in the same way as
it was achieved for 
the {HD}\,202206 \citep{Correia_etal_2005} and {HD}\,45364 \citep{Correia_etal_2009}
systems.
For each planet, the system is integrated on a regular 2D mesh of initial conditions, 
with varying semi-major axis and eccentricity, while the other parameters are 
retained at their nominal values. The solution is integrated over 10~kyr 
for each initial condition and a stability indicator is computed 
to be the variation in the measured mean motion over   two consecutive 
5~kyr intervals of time.  The results,
for different values of the inclination of the system ($i=90^\circ, 20^\circ, 10^\circ$),
 are reported in color in Fig.\,\ref{F4},
 where ``red'' represents the strongly chaotic trajectories, and ``dark blue'' 
the extremely stable ones. 
In all these plots, the 3:1 resonant island is clearly highlighted, and appears to be 
very stable until the inclination reaches $i=10^\circ$, for which most of the 
resonant island is destabilized. Indeed, a numerical integration of 
the fitted solution for 10 degrees inclination led to a disruption of the system in 
584 Myr.

It is  remarkable that, as for  {HD}\,45364 \citep{Correia_etal_2009},
there is perfect coincidence between
the stable 3:1 resonant islands, and curves of minimal $rms$
obtained by comparison with the observations. Moreover, 
for our preferred solutions, with inclination ranging from 90 to 15 degrees, 
the 
resonant island extends to large values of the $rms$ of about 7 m/s, 
which reinforces the confidence that the present system 
is in a 3:1 resonant state.  

The present dynamical analysis 
also indicates that this
3:1 mean motion resonance  is stable over Gyrs timescale.
We have as well tested directly this stability by performing a numerical integration
of the   systems $\Sa$  and $\Sb$ (Tab. \ref{T1}) 
over 5~Gyr using the symplectic integrator SABAC4 of
\citet{Laskar_Robutel_2001} with a step size of 0.02~years. 
Both solutions $\Sa,\Sb$ remain stable over 5 Gyr, although their orbital elements present 
large variations 
because of the strong gravitational interactions between the two planets.
For the nominal solution $\Sa$,
the eccentricity of the inner planet oscillates within $ 0.115 < e_b < 0.301 $, while for
the outer planet, we have $ 0.0165 < e_c < 0.143 $.
The    secular variations in the orbital parameters are mostly driven by
the rapid secular frequency $g_1$, of period $ 2 \pi / g_1 \approx 754 $~yr
(Table\,\ref{T3}).
These secular variations in the orbital elements occur much more rapidly than in
our Solar System, which may enable them to be detected directly from
observations.

\section{Additional constraints}
\begin{figure}
  \centering
    \includegraphics[width=8.5cm]{\figpath 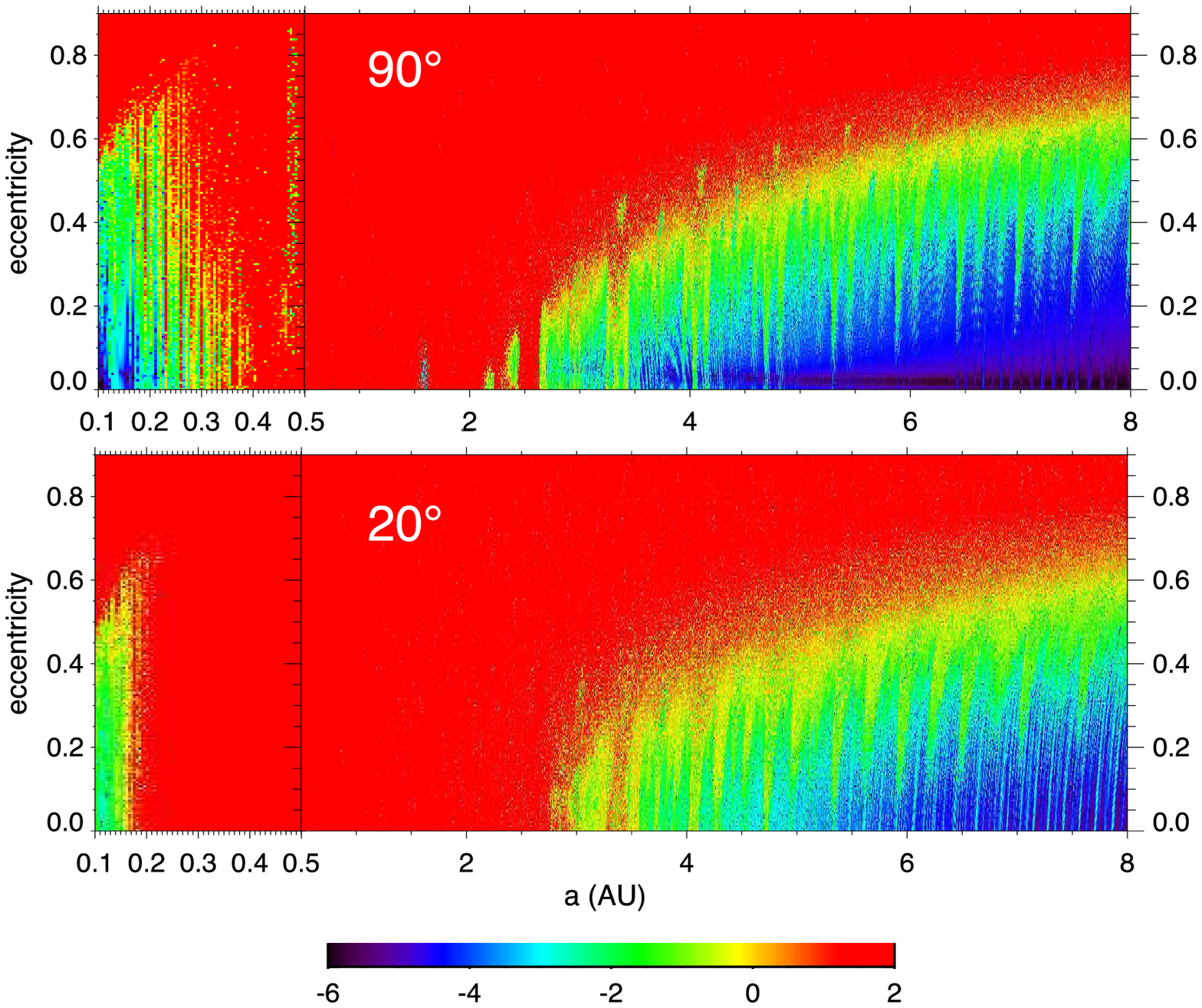} 
  \caption{Possible location of additional planet in the HD 60532 system.
  For 90 (top) and 20 degrees (bottom) inclination, the stability of a small mass 
  particle in the HD 60532 system is analyzed, for various semi-major axis and 
  eccentricity, and for $K=10^{-6}$m/s. The stable zones  where additional planets 
  could be found are the dark blue regions.
    \label{F3}}   
\end{figure}

The stability analysis summarized in Figure \ref{F4} shows, 
as in HD 45364 \citep{Correia_etal_2009}, a very good agreement between 
the 3:1 resonant islands and the $rms$ contour curves. We can thus assume that 
the dynamics of the two known planets is not much disturbed by the presence
of an additional large planet close by.

We have tested the possibility of an additional planet in the systems $\Sa$ and $\Sb$  
by including  a third planet d, 
with varying   semi major axis and eccentricity  over a large range, 
and performing   stability analysis (Fig. \ref{F3}). The test was done  for a fixed 
$K$ value ; first with an asteroid size object ($K=10^{-6}$m/s)
(Fig.\ref{F3}), and then for an Earth-size object ($K=0.1$m/s) without noticeable change
in the dynamics.

From this analysis, one can see (Fig. \ref{F3}), that for $\Sa$  ($i=90^\circ$),
the zone of instability extends from about 0.2 AU to about 4 AU,
 while 
for $\Sb$ ($i=20^\circ$), no planet can be stable from 0.1 AU to about 6 AU.
The eventual discovery of an additional planet in this range would thus 
constrain very much the inclination of the system with respect to the 
plane of view,  and thus the planetary masses.
At $i=90^\circ$, the libration period of the 3:1 resonant argument is  39.93 yr,
but this period depends strongly on the planetary masses, and thus on the 
inclination  $i$ of the system
(Tab. \ref{T4}). We have  computed the libration period $P_\theta$ 
for the best fit obtained at various inclinations, $i=90,30,20,15,10$
degrees. At $i=20^\circ$,  we have $P_\theta=23.17$ yr.
As the amplitude of the libration is large, we can expect that 
over the next decade, this resonant libration period will be 
constraint, allowing a good determination of the 
inclination and planetary masses. Indeed, if we extrapolate the differences 
of radial velocity, for various inclination hypothesis 
($i=30,20,15,10$ degrees), compared with the nominal solution $\Sa$, 
it appears clearly that the inclination of the system will be determined 
within a few years with the HARPS data (Fig.\ref{F4}).

\begin{table}
 \caption{Change of the mass 
 factor ($1/\sin i $) and libration period $P_\theta$   with inclination ($i$). 
 \label{T4}} 
 \begin{center}
 \begin{tabular}{crrrrr}
 \hline\hline
$i$    & $90^\circ  $ & $30^\circ  $& $20^\circ  $& $15^\circ  $& $10^\circ  $ \\
 \hline
$1/\sin i$ & 1.00  &      2.00  &      2.92  &      3.86  &      5.76    \\
 $P_\theta$ (yr) & 39.93  &     28.88  &     23.17  &     19.57  &     15.35    \\ \hline
\end{tabular}
\end{center}
\end{table}


\section{Discussion and conclusion}
\begin{figure}
  \centering
    \includegraphics[width=8.5cm]{\figpath 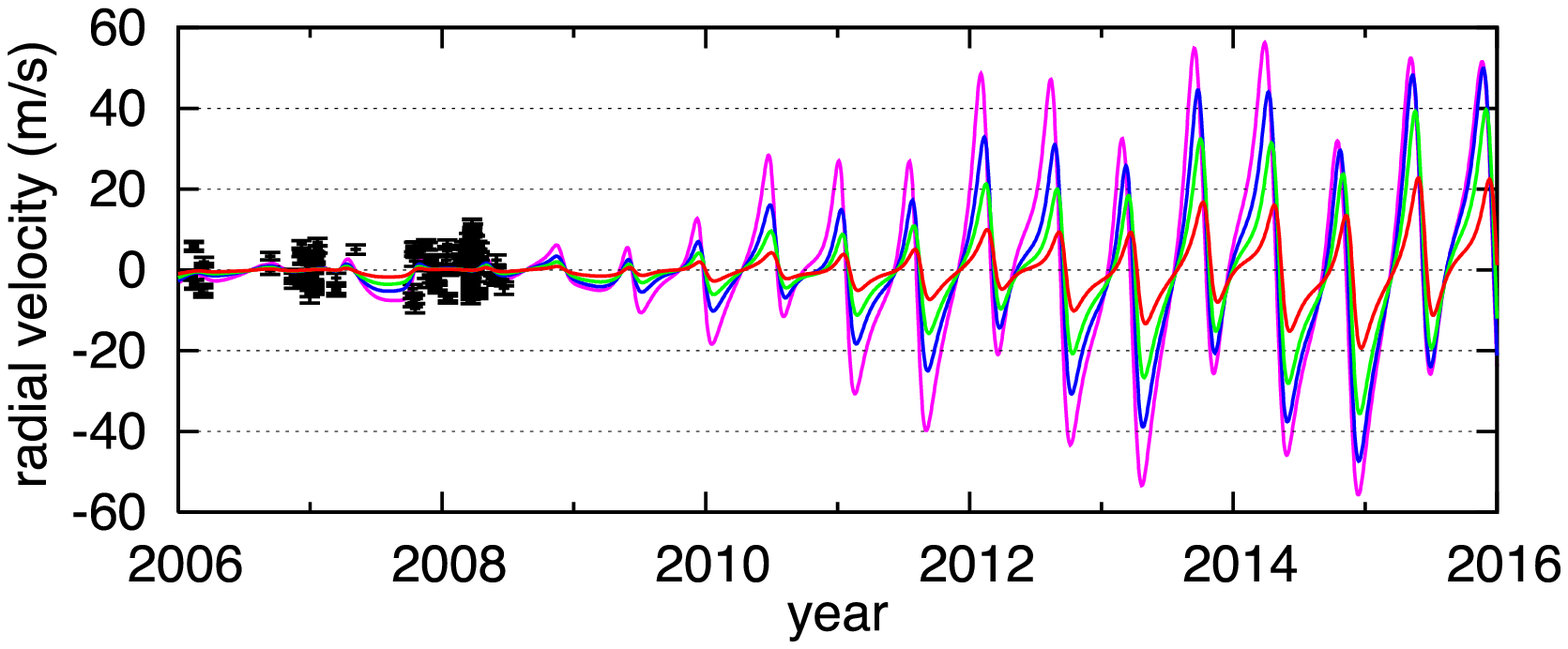} 
  \caption{Radial velocity signature of the inclination ($i$).
  The difference with respect to the nominal solution (at $i=90^\circ$), 
  are plotted  over time, for $i=30^\circ$ (red), $20^\circ$ (green),
  $15^\circ$ (blue), and $10^\circ$ (violet). The available HARPS residual data are also 
  plotted with their error bars.
    \label{F4}}   
\end{figure}

We have re-analyzed the dynamics of the two planets system 
{HD}\,60532, first reported in DES08. 
Contrary to the conclusions of this previous work, we believe 
that the global dynamical study presented here 
(Fig.\ref{F4}) allows to conclude unambiguously that the system is
in a 3:1 mean motion resonance. 

Moreover, due to the strong dynamical interaction present in this 
system, we could also fit the inclination 
which led to a most probable value of 20 degrees, while our 
dynamical analysis set up a stability constraints 
to $15^\circ \leq i \leq 90^\circ$. 
The confidence in the presence of the 3:1 resonance is enhanced by 
the fact that the system remains in resonance, for all the fits 
that we performed, from 90 to 15 degrees inclination.
In the nominal solution ($\Sa$, Tab.\ref{T1}), 
the resonant angle $ \theta  =    \lambda_b - 3 \lambda_c + 2\omega_b $ is in libration
around $ 180^\circ $, with a  libration period of 39.93~years
with an associated amplitude of $38\fdg 14$, although the 
main oscillation of this argument  is driven by the secular 
term $g_1-g_2$ with an associated amplitude of $63\fdg 65$, 
the total amplitude of the libration being $123\fdg 16$, 
due to additional harmonics.

The planet-planet interactions in this system are large, due to the presence 
of this 3:1 resonance. The  dependence of the libration period with 
the mass factor (Tab. \ref{T4}) 
should should allow to determine the
inclination of the orbital planes and thus  the  masses
of both planets within the next decade (Fig.\ref{F4}).


\begin{acknowledgements}
We acknowledge support from the 
Funda\c{c}\~{a}o Calouste Gulbenkian (Portugal) and French PNP-CNRS.
This work was elaborated during a stay of the authors at Geneva Observatory,
where discussions with M. Mayor and  S. Udry were very helpful.
\end{acknowledgements}

\bibliographystyle{aa}
\bibliography{correia}

\end{document}